# 99% efficiency in collecting photons from a single emitter

Xue-Wen Chen[1], Stephan Götzinger[1] and Vahid Sandoghdar[2*]

[1]Laboratory of Physical Chemistry, ETH Zürich, CH-8093 Zürich, Switzerland
[2]New address: Max Planck Institute for the Science of Light, D-91058 Erlangen, Germany
*Corresponding author: vahid.sandoghdar@mpl.mpg.de

In *Nature Photonics* **5**, 166 (2011), we reported on a planar dielectric antenna that achieved 96% efficiency in collecting the photons emitted by a single molecule. In that work the transition dipole moment of the molecule was set perpendicular to the antenna plane. Here, we present an extension of that scheme that reaches collection efficiencies beyond 99% for emitters with arbitrarily oriented dipole moments. Our work opens important doors in a wide range of contexts including quantum optics, quantum metrology, nano-analytics, and biophysics. In particular, we provide antenna parameters to realize ultrabright single-photon sources in high-index materials such as semiconductor quantum dots and color centers in diamond, as well as sensitive detection of single molecules in nanofluidic devices.

Detection, microscopy and spectroscopy of single optical emitters such as organic molecules, semiconductor quantum dots, and color centers have enabled a vast range of studies in the past two decades [1, 2]. One of the outstanding impacts of this research field has been the demonstration of single-photon sources based on the antibunched nature of the radiation from a single emitter [2-7]. Another important widespread application has been in single-molecule biophysics [8, 9]. Both of these lines of research are limited by the intrinsically weak radiation of a single emitter and would, therefore, massively benefit from more efficient collection strategies. Indeed, over the years many groups have addressed this issue in different ways [10-12]. Recently, we presented a theoretical and experimental breakthrough in collection efficiency, reaching 96% for emitters with transition dipole moments perpendicular to the surface of a planar dielectric antenna [13]. In this Letter, we extend that design concept to go beyond 99% collection efficiency without any restrictions on the orientation of the transition dipole moment.

The proposed antenna configuration is sketched in the inset of Fig. 1(a), where a metallic mirror is placed on a dielectric structure. A single emitter with an arbitrary dipole moment orientation is embedded in a thin film of refractive index $n_2$, which we will call medium 2. This medium is sandwiched between media 1 and 3 with indices of refraction $n_1$ and $n_3$ ($n_1 > n_2 > n_3$), respectively. Without the upper mirror, the lower medium 1 funnels the photons that couple to the quasi-waveguide modes of medium 2 into a cone with an opening angle within the reach of commercial microscope objectives [13]. The shortcoming of this configuration, however, is that emitters with dipole moments oriented in the plane of the antenna lose about 15% of their power to the outside medium on the upper side. Adding a metallic layer remedies this difficulty, but it can also introduce absorptive losses as well as leakage to surface plasmon polariton (SPP) modes. In what follows, we discuss the performance of this new metallo-dielectric antenna.

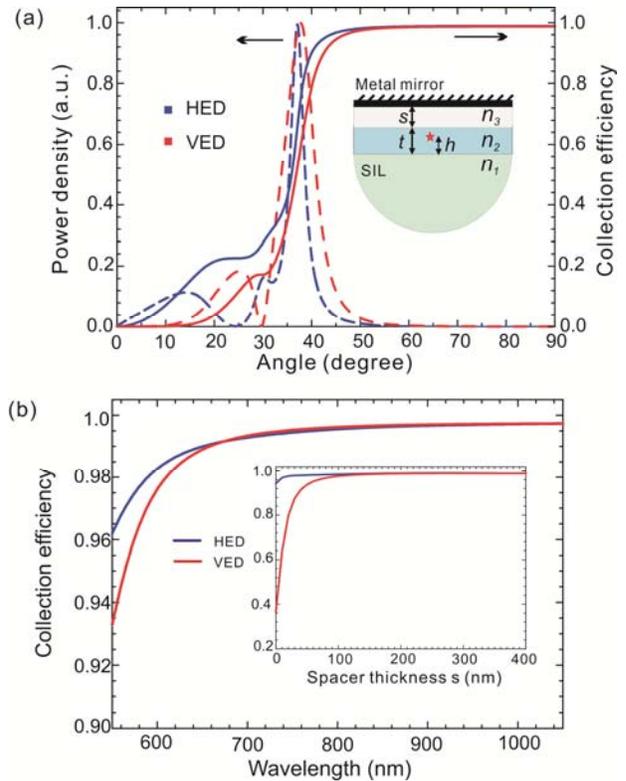

Fig. 1 (Color online) (a) Power density (left axis) and collection efficiency (right axis) versus emission angle at a wavelength of 637 nm. The inset shows the schematics of the device. (b) Collection efficiency as a function of wavelength. The inset plots the collection efficiency versus spacer thickness $s$ at 637 nm; SIL: solid-immersion lens. For both (a) and (b) we used $n_1 = 2.2$, $n_2 = 1.5$, $n_3 = 1.0$, $t = 350$ nm, $h = 200$ nm and $s = 200$ nm.

As an example of a system of current interest, we first consider diamond or colloidal semiconductor nanocrystals embedded in a thin polymer film of thickness $t$ and refractive index $n_2 = 1.5$. We take the underlying substrate to be a thick medium of zirconia with $n_1 = 2.2$, which is formed as a solid immersion lens (SIL). The simplest spacer layer could be air with $n_3 = 1$ or silica aerogel with an index close to 1 [14]. Finally, we place a gold mirror on top. To evaluate the collection efficiency and the angular distribution of the emission, we employed the plane wave expansion method [15] and calculated the total emission power, the power released to the substrate, and its distribution using the values in Ref. [16] for the dielectric function of gold.

Figure 1(a) displays the calculated angular distribution of the emission at $\lambda = 637$ nm corresponding to the zero-phonon line of the nitrogen-vacancy color center in diamond. Here, we took $t = 350$ nm, $s = 200$ nm, and $h = 200$ nm, where $h$ denotes the distance of the emitter to the interface between media 1 and 2. The dashed red and blue traces depict the angular dependence of the power density for vertical electric dipoles (VEDs) and horizontal electric dipoles (HEDs), respectively. Both traces show well-behaved emission patterns that allow easy collection and manipulation. The solid red and blue curves show the collection efficiency as a function of the collection angle. We find the remarkable result that 99% of the total emitted photons can be captured within an angle of 55°. This cut-off angle can be further reduced to 43° ($\sin^{-1}(n_2/n_1)$) by increasing $h$.

Our design requires the condition $n_1 > n_2 > n_3$ and that $h$ be kept about or above half of an effective wavelength ($\lambda/2n_2$). The index of refraction $n_3$ of the spacer layer has to be kept low to minimize the coupling to SPP. Moreover, by adjusting the emitter position and spacer thickness, the coupling to SPP can be reduced to below 1%.

Figure 1(b) plots the dependence of the collection efficiency on the emission wavelength. Interestingly, efficiencies beyond 99% can be achieved in a broad spectral range for both VED and HED. At shorter wavelengths the absorption of gold begins to reduce the efficiency. The inset in Fig. 1(b) examines the influence of the spacer thickness $s$ on the collection efficiency at $\lambda=637$nm. In the case of a VED, the collection efficiency drops rapidly for $s<100$nm because $k_{sp}$ increases due to the change of the refractive index of medium 3 from 1 to 1.5. Here, $k_{sp} = k_0[\varepsilon_{au}\varepsilon/(\varepsilon_{au}+\varepsilon)]^{1/2}$ is the wave vector of the SPP mode and $\varepsilon_{au}$ and $\varepsilon$ are the dielectric constants of gold and medium 3, respectively. The coupling of a VED to SPP is evanescent and is, therefore, significantly suppressed for smaller $k_{sp}$ and at larger distances from the metal surface. The coupling to SPP from an HED is much weaker because the generated electric fields have smaller vertical components. Thus, the influence of the spacer layer is less and the coupling to SPPs can be suppressed by simply increasing the emitter-metal distance even for $s = 0$.

The example in Fig. 1 is only one realization for a certain choice of refractive indices, but the proposed device configuration also works well for other materials as long as the above-mentioned design rules are satisfied. For a given active medium 2, we need $n_1 > n_2$ to have near-unity collection with an angle determined by $\sin^{-1}(n_2/n_1)$. However, the achievable collection efficiency mainly depends on the spacer index $n_3$ and its thickness. Figure 2 shows the collection efficiency as a function of the ratio $n_2/n_3$ for various $n_2$ values and $n_1/n_2=1.2$, while the other parameters are kept the same as in Fig. 1(a). The upper (a) and lower (b) plots report on the cases of HED and VED, respectively. For the former, the collection efficiency is about 98% or above for any choices of $n_2/n_3 \geq 1$. For a VED, the efficiency depends on both $n_2/n_3$ and $n_2$. But as $n_2/n_3$ exceeds a certain value, the collection efficiency can be kept more than 99% for any choice of $n_2$. The inset shows this lower bound as a function of the spacer thickness. One sees that the threshold value of $n_2/n_3$ decreases as the spacer thickness is increased from 200 nm to 800 nm.

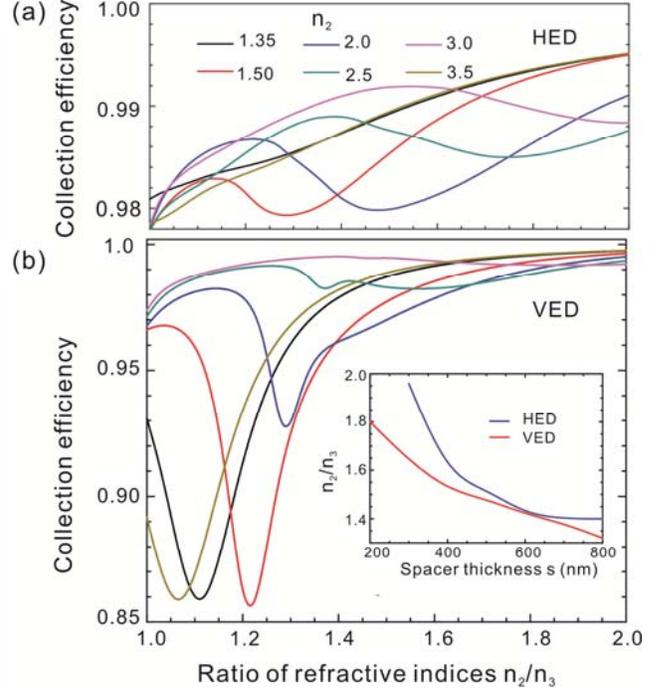

Fig. 2. (Color online) Collection efficiencies of horizontally (a, HED) and vertically (b, VED) emitting dipoles versus the ratio $n_2/n_3$ for various $n_2$ values while $n_1/n_2$ is fixed to 1.2. The other parameters are the same as in Fig. 1(a). The inset depicts the lower bound on $n_2/n_3$ on the spacer thickness if a collection efficiency over 0.99 is to be achieved.

The limited availability of high-index materials poses a practical problem in meeting the condition $n_1/n_2>1$ for media such as diamond, GaAs or other semiconductors, which have been of interest as hosts for single-photon sources [6,7, 10-12,17]. To get around this hurdle, one can introduce a thin buffer layer with a low refractive index between media 2 and 1. Intuitively speaking, this reduces the effective refractive index of medium 2. The inset in Fig. 3(a) sketches an example of a quantum dot embedded in a GaAs membrane with $n_2=3.5$, surrounded by medium 3 with 400 nm thickness and a low refractive index. The main part of Fig. 3(a) displays a set of parameters that allow collection efficiencies as large as 99% at $\lambda=900$ nm for a substrate with refractive index $n_1 =3.5$. In fact, the high index of medium 1 ensures small exit angles that can be easily captured using lens systems with low numerical aperture.

We note that although the epitaxic growth of crystalline matrices is restricted to a limited class of substrates, it is possible to transfer a properly grown thin film onto other materials [18]. Furthermore, membranes can be fabricated as air bridge by a combination of lithography and dry and wet etching [19]. Efficient extraction of photons from high-index materials has been an outstanding problem of important technological implications. The great advantage of our approach over the existing strategies based on roughened interfaces [20] is the access to well-defined optical modes.

Efficient collection of fluorescence is also highly desirable in single-molecule detection and analytics [21]. Since photobleaching restricts the total number of photons emitted by a fluorophore, it is crucial to collect as much of its emission as possible to improve the shot-noise limited signal-to-noise ratio. A particularly important but challenging class of applications takes place in the liquid phase, where the dipole moments of the fluorophores rotate rapidly. To achieve directed emission in a small solid angle from such emitters, we propose to confine the liquid of interest in a nanofluidic slit.

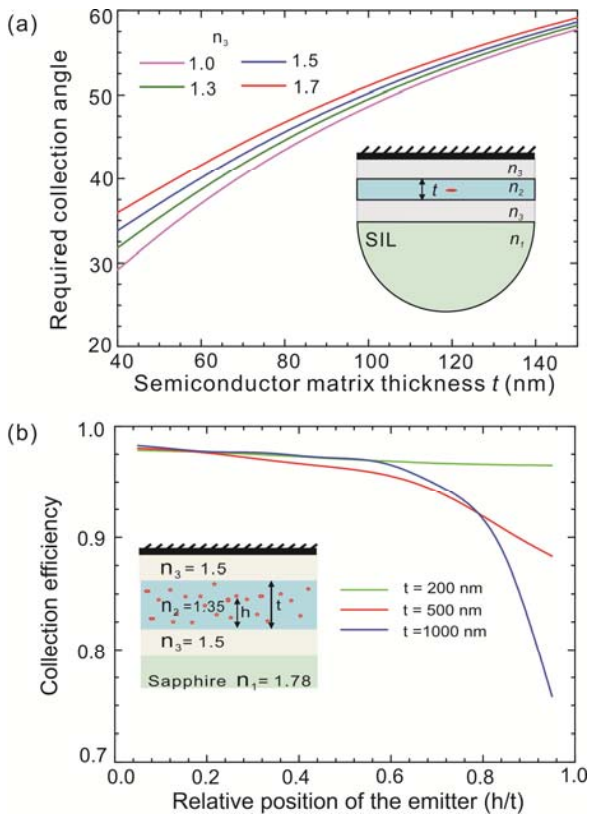

Fig. 3. (Color online) (a) The collection angle required for achieving more than 99% efficiency versus the thickness of a semiconductor matrix ($n_2 = 3.5$) for $n_1 = 3.5$ and different $n_3$ values at $\lambda = 900$ nm. The inset displays the antenna arrangement. (b) Collection efficiency versus relative position of a randomly oriented fluorophore flowing in channels of three different thicknesses.

The conceptual device structure is sketched in the inset of Fig. 3(b). A nanofluidic channel is created between two media with $n_3=1.5$, and fluorophores flow in the fluidic region of $n_2 = 1.35$. The top medium is coated with a layer of gold while the bottom one is placed on a sapphire substrate. Similar structures have been successfully used for a variety of experiments [21, 22] and can be fabricated even in a suspended geometry. The color-coded lines in Fig. 3(b) show the fluorescence collection efficiency as a function of the relative position of a randomly-oriented fluorophore at $\lambda=650$ nm for three different channel sizes, 200 nm, 500 nm and 1000 nm respectively. For the channel size of 200 nm, the collection efficiency is always above 94% regardless of the position of the fluorophore. For channel depths of 500 nm and 1000 nm, collection efficiencies are also well above 90% for h/t<0.8. Thus, the proposed device is very effective for collecting fluorescence from a nanofluidic device and lab-on-chip arrangements.

We have proposed a robust metallo-dielectric planar antenna with near-unity efficiency for collecting photons emitted by randomly oriented emitters. Some of the outstanding features of our antenna design are that its operation is broadband, it does not require sophisticated lateral nanofabrication, and it is compatible with essentially all materials. Our results have immediate applications in quantum optics, quantum metrology [23, 24], spectroscopy, photonics, and nano-analytics. In particular, the realization of single-photon devices with efficiency over 99% will provide non-classical light with sub-shot noise level, enabling quantum microscopy and spectroscopy.

We acknowledge the financial support of ETH Zurich and the Max Planck Society.